  \documentstyle[12pt]{article}
  
\catcode`@=11
\def\secteqno{\@addtoreset{equation}{section}%
\def\theequation{\thesection.\arabic{equation}}}
\catcode`@=12
\secteqno
\newcommand{\be}{\begin{equation}}
\newcommand{\ee}{\end{equation}}
\newcommand{\bea}{\begin{eqnarray}}
\newcommand{\eea}{\end{eqnarray}}
\newcommand{\bref}[1]{(\ref{#1})}
\newcommand{\nn}{\nonumber}

\newcommand{\osp}{osp$(1|32)~$}
\newcommand{\ads}{AdS$_5\times$S$^5~$}
\newcommand{\sads}{super-AdS$_5\times$S$^5~$}
\newcommand{\scn}{superconformal~}
\newcommand{\Jis}{Jacobi identities }\newcommand{\Ji}{Jacobi identity }
\newcommand{\A}{\alpha} \newcommand{\B}{\beta} 
\newcommand{\gam}{\gamma}\newcommand{\G}{\gamma} 
\newcommand{\Gam}{\Gamma} 
\newcommand{\D}{\delta} 
\newcommand{\ep}{\epsilon}

           \newcommand{\s}{\sigma}

\newcommand{\h}{\eta}           
           
\newcommand{\W}{\Omega}

\def\6{\partial}
\def\7{\tilde}
\def\8{\hat}

\def\hz{\hat {0}}

\def\1{1\hspace{-1.5mm}1}

\def\CA{{\cal A}}\def\CB{{\cal B}}\def\CC{{\cal C}}\def\CD{{\cal D}}
\def\CE{{\cal E}}\def\CF{{\cal F}}\def\CG{{\cal G}}
\def\CQ{{\cal Q}}
\def\CZ{{\cal Z}}
\def\t{\hat}
\def\bQ{{\bf Q}}\def\bS{{\bf S}}
\def\hI{{\hat I}}\def\hJ{{\hat J}}\def\hK{{\hat K}}\def\hz{{\hat 0}}
\def\vs{\vskip 3mm}
\begin{document}
\thispagestyle{empty}
\vfill 
\hfill January 14, 2003\par
\hfill KEK-TH-862\null\par
\hfill Toho-CP-0373\null\par
\vskip 20mm
\begin{center}
{\Large\bf osp(1$|$32) and
Extensions of super-AdS$_5\times $S$^5$ algebra \\
 }\par
{\Large\bf }\par
\vskip 6mm
\medskip
\vskip 10mm
{\large 
Kiyoshi~Kamimura~and~
Makoto\ Sakaguchi$^\dagger$ }\par
\medskip
{\it 
 Department of Physics, Toho University, Funabashi, 274-8510, Japan}\\
{\it 
$~^\dagger$Theory Division,\ 
High Energy Accelerator Research Organization (KEK),\\
 Tsukuba,\ Ibaraki,\ 305-0801, Japan }\\
\medskip\medskip
{\small\sf E-mails:\  kamimura@ph.sci.toho-u.ac.jp,
Makoto.Sakaguchi@kek.jp} 
\medskip
\end{center}
\vskip 10mm

\begin{abstract}

The \sads and the four-dimensional $\cal N$=4 superconformal 
algebras play important roles in superstring theories. 
It is often discussed the roles of the \osp algebra as a maximal 
extension of the superalgebras in flat background. 
In this paper we show that  the su(2,2$|$4), the super-AdS$_5\times S^5$ 
algebra or the superconformal algebra, is not a restriction of the osp(1$|$32)
though the bosonic part of the former is a subgroup of the latter. 
There exist only two types of u(1) extension of the \sads algebra if 
the bosonic \ads covariance is imposed. 
Possible significance of the results is also discussed briefly.

\end{abstract}

\par\par
\newpage
\setcounter{page}{1}
\parskip=7pt
\section{Introduction}\par
\indent

All possible extensions of the super-Poincare algebra
correspond to branes that can exist on the flat background \cite{T}.
The maximal extension of the super-Poincare algebra in 10 dimensions
contains 32 supercharges
and 528 bosonic generators (in addition to Lorentz generators). 
It is the osp(1$|$32) algebra and the symmetries of IIA, IIB, M and F 
theories have been examined systematically \cite{AVP}.\footnote{
Superalgebras
with a larger number of supercharges than 32
have been discussed in \cite{64}.}

A D-brane action consists of a Dirac-Born-Infeld (DBI) action
and a Wess-Zumino (WZ) action.
The WZ action is needed for the total action to possess $\kappa$-symmetry,
which allows to project out half of the world-volume fermions
and matches bosonic and fermionic degrees of freedom on the 
world-volume\cite{AETW}.
It was shown\cite{dAGIT} that 
commutation relations among two of the Noether charges
do not close but include topological brane charges, 
because the WZ action is quasi-superinvariant. 
In other words, the super-Poincare algebra is modified
to include topological brane charges and becomes
the extended super-Poincare algebra with brane charges\footnote{
The brane charge is not a center of the extended super-Poincare algebra,
because they have non-trivial commutation relations with Lorentz generators.
These brane charges are ideal of the extended super-Poincare algebra,
and the original algebra is represented as a coset,
(the extended super-Poincare algebra)/ideal.
}.
The half of the 32 supercharges are those for Nambu-Goldstone (NG) fermions
associated with supersymmetries broken by the brane.
In fact, for a given brane configuration
with the brane tension being equal to the brane charge,
anti-commutation relation of two
of supercharges turns out to be proportional to a projection
operator which projects out half of the supersymmetries
on the vacuum.

These brane backgrounds are constructed as solutions of supergravity theories.
Constructing  Killing vectors and spinors in a brane background,
one can show that the super-isometry algebra contains a smaller number of
supercharges than 32.
In order to recover (super)symmetries broken by the brane,
one examines small fluctuations around this solution.
The broken symmetries are recovered in terms of NG fields
and the resulting superalgebra contains 32 supercharges and brane charges.

For a curved background, 
D-brane actions are shown \cite{BT} to be $\kappa$-symmetric
if the background satisfies the field equations
of the target space supergravity.
Since the AdS$_5\times S^5$ background is a solution of type-IIB supergravity,
one can construct $\kappa$-symmetric D-brane action on it.
Such D-brane actions have been examined in \cite{MT:D3,CGMVP,ST}.
The amount of unbroken supersymmetries are determined by the balance between
the $\kappa$-symmetry projection and Killing spinor equations.

One expects that the WZ action for such a brane produces 
brane charges in the superalgebra su(2,2$|$4), which is the 
super-isometry algebra of the AdS$_5\times S^5$ background,
and the maximal extension is osp(1$|$32),
referring the classification by Nahm \cite{Nahm}.
The anti-commutator between two of supercharges are examined 
in \cite{SY} for a matrix theory on the eleven-dimensional pp-wave and 
in \cite{HS} for a superstring action on the AdS$_5\times S^5$ background,
and are shown to include brane charges.
However, the full supersymmetry algebra has not been derived yet 
because of the complexity.
To derive the full supersymmetry algebra,
one may try to examine small fluctuations around a D-brane solution
of a supergravity theory.
But supergravity solutions for D$p$-branes in the AdS$_5\times S^5$ background
have not been known well yet.

In  order to achieve this, we take another route in this paper.
We examine the osp(1$|$32) and relations to
the super-AdS$_5\times S^5$ algebra, su(2,2$|$4).
The bosonic AdS$_5\times S^5$ algebra,
so(4,2)$\times$so(6), is a subalgebra of sp(32).
However, su(2,2$|$4) is not a subalgebra of osp(1$|$32)
because commutation relations, for example $\{Q,Q \}$, differ each other.
Algebraically, 
the osp(1$|$32) is not an extension of su(2,2$|$4) because
sp(32) is not an extension of so(4,2)$\times$so(6).

In this paper
we find that the su(2,2$|$4) is not a restriction of the osp(1$|$32)
in the following sense. Expressing the osp(1$|$32) algebra in an AdS$_5\times 
S^5$ covariant basis and restricting the bosonic generators of sp(32) to 
those of so(4,2)$\times$so(6),
{ the resulting commutation relations have forms of \sads algebra 
apart from some sign differences.  
We find that they 
are neither those for the 
su(2,2$|$4) nor a consistent superalgebra. }
This is essentially because the Fierz identity for the one hand does not
imply that for the other hand.
The same argument is true for the four-dimensional ${\cal N}=4$ superconformal 
algebra.

In the reverse, we examine possible generalizations
of the super-AdS$_5\times S^5$ algebra.
When we consider some (brane) solutions in the \sads background 
they break also manifest $AdS_5$ and/or $S^5$ covariance.
However it is a spontaneous symmetry breaking
in the presence of a particular solution. 
The \ads covariance of the theory would be still maintained. 
Thus,
we impose the assumption that the bosonic AdS$_5\times S^5$ algebra
is a subalgebra of the bosonic part of the superalgebra with brane charges,
and find that only two trivial types of u(1) extension of the
super-AdS$_5\times S^5$ algebra are allowed.

This paper is organized as follows.
In the next section, 
we express  the osp(1$|$32) algebra in an AdS$_5\times S^5$ covariant basis.
In section 3, we show that the super-AdS$_5\times S^5$ algebra is not a 
restriction of the osp(1$|$32) algebra.
In section 4, we find that only two types of u(1) extension of the
super-AdS$_5\times S^5$ algebra are allowed,
under the assumption that bosonic AdS$_5\times S^5$ algebra is a subalgebra of
the bosonic part of the superalgebra with brane charges.
The section 5 is devoted to a summary and discussions on possible significance.
{The uniqueness of the osp(1$|$32) algebra is presented in appendix A.}
In appendix B, complementary to the section 3,
we show that the four-dimensional $\cal N$=4 superconformal algebra is not
a restriction of osp(1$|$32),
by clarifying the relation between generators of 
the super-AdS$_5\times S^5$ algebra
and those of the four-dimensional $\cal N$=4 superconformal algebra.

\section{ \osp }\par
\indent

The  \osp is a maximally extended supersymmetry algebra whose
generators are 32 supercharges $\CQ_{\CA}$ ($\CA=1,2,...,32)$ and 528 
bosonic ${\CZ}_{{\CA}{\CB}}$. 
The algebra is \cite{AVP}
\bea
\{\CQ_{\CA},\CQ_{\CB}\}&=&{\CZ}_{{\CA}{\CB}},
\label{ospqq}\\
{[}\CQ_{\CA},{\CZ}_{{\CB}{\CC}}]&=&\CQ_{({\CB}}\W_{{\CA}{\CC})},
\label{ospqz}\\
{[}{\CZ}_{{\CA}{\CB}},{\CZ}_{\CC{\CD}}]&=&
\W_{{\CA}({\CC}}{\CZ}_{{\CD}){\CB}}~+~\W_{{\CB}
({\CC}}{\CZ}_{{\CD}){\CA}}
\label{ospzz}
\eea
where ${\CZ}_{{\CA}{\CB}}$ is $32\times 32$ symmetric matrix and 
$\W_{{\CA}{\CB}}$ is an anti-symmetric invertible symplectic metric.
The \Jis hold {identically} for any anti-symmetric $\W$. 
It is shown in appendix A 
that any maximally extended supersymmetric algebra is 
expressed in the form of \bref{ospqq}-\bref{ospzz} by redefinitions. 

The supercharge $\CQ_{\CA}$'s are 32 complex generators and are
subject to a Majorana condition reducing the number of independent 
degrees of freedom to half,
\bea
{\CQ^\dagger}^{\CA}&=&\CQ_C{{B}^{{\CC}{\CA}}}~=~{{B^t}^{{\CA}{\CC}}}\CQ_C
\label{conjq}\eea
for some non-singular matrix $B$ satisfying 
\bea
{\W^\dagger}^{{\CA}{\CB}}&=&({B^{t}}\W{B})^{{\CA}{\CB}},~~~~~
{\rm and}~~~~~B^{\CA\CB}~B^*_{\CB\CC}~=~{\D^\CA}_{\CC}.
\label{conjW}
\eea
The  \osp algebra  \bref{ospqq}-\bref{ospzz}
remains unchanged under the conjugation if $\CZ$'s satisfy 
\bea
{{\CZ}^\dagger}^{{\CA}{\CB}}&=&
( B^t{\CZ}B)^{{\CA}{\CB}}.
\label{Zdagger}
\eea
in addition to \bref{conjq} and \bref{conjW}.

In this paper we are interested in relations among the 
 \osp, \sads and \scn algebras 
we represent the $32$ component supercharge $\CQ_{\CA}$ as a
$SO(4,1)\times SO(5)\times SO(2,1)$ spinor, \cite{MT}
\bea
\CQ_\CA&=&Q_{\A\A' A},~~~~~(\A=1,2,3,4,~\A'=1,2,3,4,~A=1,2),
\eea
where $\CA$ is a collective index of $(\A\A'A)$. 
The gamma matrices and charge conjugation for SO(4,1) satisfy, 
$(a=0,1,2,3,4)$,
\bea
\{\G^a,{\G}^b\}&=&2\h^{ab}=2~diag(-++++),~~~~~~~~
{\G}^0{\G^a}^\dagger\G^0~=~{\G^a}
\nn\\
C^t&=&-C,~~~~~~~~C^\dagger C~=~1,~~~~~~~~C\G^aC^{-1}~=~({\G^a})^t,
\nn\\
&&~(C\G^a)^t~=~-~(C\G^a) ~~~~~~~(C\G^{ab})^t~=~(C\G^{ab}),
\eea
those for SO(5) are,  $(a'=1',2',3',4',5')$,
\bea
\{\G^{a'},\G^{b'}\}&=&2\h^{a'b'}=2~diag(+++++),~~~~~~~~
{\G^{a'}}^\dagger~=~{\G^{a'}}
\nn\\
{C'}^t&=&-C',~~~~~~~~{C'}^\dagger C'~=~1,~~~~~~~~
C'{\G^{a'}}{C'}^{-1}~=~({\G^{a'}})^t,
\nn\\
&&~(C'\G^{a'})^t~=~-~(C'\G^{a'}) ~~~~~~~(C'\G^{a'b'})^t~=~(C'\G^{a'b'}).
\eea
For SO(2,1) they are, $(j=\hat 0,\hat 1,\hat 2)$,
\bea
{\{}\rho^i,\rho^j\}&=&2\h^{ij}~=~2~diag(-,++),~~~~~~~~
{\rho^{\hz}}{\rho^{j}}^\dagger{\rho^{\hz}}~=~{\rho^{j}},
\nn\\
c&=&-c^{~t},~~~~~~~~{c}^\dagger{c}~=~1,~~~~~~~~c\rho^jc^{-1}~=~-({\rho^j})^t
\nn\\
&& ~(c\rho^j)^t~=~(c\rho^j).
\eea
The real forms of $c$ and $\rho$'s we use here are 
\bea
c&=&i\tau_2,~~~~\rho^j~=~(i\tau_2,\tau_1,-\tau_3),~~~~
c\rho^j~=~(-1,\tau_3,\tau_1).
\eea

Using these notations the symplectic metric $\W$ is taken to be 
proportional to a charge conjugation $\CC$,
\bea
\W&=&s~\CC,~~~~~~~~s^*~=~s,
\label{Omega}
\eea
where $\CC$ is a total charge conjugation defined as
\bea
\CC&\equiv&c~C~C',~~~~\CC^t~=~-\CC,~~~~\CC^\dagger\CC~=~1.
\eea
The Majorana condition \bref{conjq} is imposed as
\bea
\CQ^\dagger&=& \CQ~B,~~~~~~~B~=~-e^{i\chi}~ (\rho^0 \G^0)~\CC^{-1}.
\label{MconjQ}
\eea
$e^{i\chi}$ is a phase ambiguity and taken to be $1$ in the following.
The explicit form of \bref{MconjQ} in component form is one of \cite{MT}
\bea
{Q^\dagger}^{\A\A'A}&=&
Q_{\B\B'B}~{\D}^{BA}{{(\G^0C^{-1})}^{\B\A}}
{({C'}^{-1})}^{\B'\A'}.
\label{conjugation}
\eea
The condition \bref{conjW} is verified as
\bea
({B^{t}}\W{B})&=&(\rho^0 \G^0\CC^{-1})^t~(s\CC)~(\rho^0 \G^0\CC^{-1})~=~
s~\CC^{-1}~=~{\W^\dagger}.
\eea

\vs

Using products of these gamma matrices we can construct a complete basis of 
symmetric matrices $\{(\CC\Gam^{(\hI)})_{\CA\CB}\}$, 
($\hI=1,2,...,528=\frac{32\cdot 33}{2}$),
\bea
\CC\Gam^{(\hI)}&=&(\CC\Gam^{(\hI)})^t,~~~~i.e.~~~~
\Gam^{(\hI)t}~=~-\CC\Gam^{(\hI)}\CC^{-1}.
\label{CGam}
\eea
They are
\bea\matrix{
\rho^j&\equiv &\Gam^{j}~,&~~~~&({\bf 3,1,1})\cr
i~\rho^j~\gam^{a}&\equiv &\Gam^{ja}~,&~~~~&({\bf 3,5,1})\cr
\rho^j~\gam^{a'}&\equiv &\Gam^{ja'}~,&~~~~&({\bf 3,1,5})\cr
\gam^{ab}&\equiv &\Gam^{ab}~,&~~~~&({\bf 1,10,1})\cr
\gam^{a'b'}&\equiv &\Gam^{a'b'}~,&~~~~&({\bf 1,10,1})\cr
\gam^{ab}~\gam^{c'}&\equiv&\Gam^{abc'}~,&~~~~&({\bf 1,10,5})\cr
i~\gam^{a}~\gam^{b'c'}&\equiv &\Gam^{aa'b'}~,&~~~~&({\bf 1,5,10})\cr
i~\rho^j~\gam^{a}~\gam^{a'}&\equiv &\Gam^{jaa'}~,&~~~~&({\bf 3,5,5})\cr
\rho^j~\gam^{ab}~\gam^{a'b'}&\equiv&\Gam^{jaba'b'}~,&~~~~&({\bf 3,10,10})\cr}
\label{Gsymgam}
\eea
where $({\bf a,b,c})$ indicates the decomposition under 
$SO(2,1)\times SO(4,1)\times SO(5) $ representations. 
Factor $``i"$ is introduced in \bref{Gsymgam}
for the conjugation \bref{conjugation} so that they satisfy
\bea
\rho^\hz\G^0{\Gamma^{(\hI)}}^\dagger \rho^\hz\G^0&=&-~\Gamma^{(\hI)}.
\label{Gdagger0}
\eea

$\Gamma_{(\hI)}$ is defined as inverse of  $\Gamma^{(\hI)}~,~$\footnote{
The Killing metric 
$g_{\hI\hJ}=
\frac{1}{32}tr(\Gamma_{(\hI)}\Gamma_{(\hJ)})$ 
is diagonal. The diagonal elements are either $1$ or $-1$ and 
$\Gamma_{(\hI)}=g_{\hI\hJ}\Gamma^{(\hJ)}$ is inverse of $\Gamma^{(\hI)}$. 
$~{f_{\hI\hJ\hK}}=g_{\hK\hat L}{f_{\hI\hJ}}^{\hat L}$ 
is totally anti-symmetric.}
, e.g.
\bea
(\CC\Gamma^{(ab)})=(\CC\gamma^{ab})~~~~&\to&~~~~
(\Gamma_{(ab)}\CC^{-1})=(\gamma_{ba}\CC^{-1}),
\\
(\CC\Gamma^{(ja)})=(i\CC\rho^j\gamma^{a})~~~~&\to&~~~~
(\Gamma_{(ja)}\CC^{-1})=(-i\rho_j\gamma_{a}\CC^{-1}).
\eea
They satisfy orthonormal and completeness relations
\bea
\frac{1}{32}~(\Gamma_{(\hI)}\CC^{-1})^{{\CB}\CA}~
(\CC\Gamma^{(\hJ)})_{{\CA}{\CB}}&=&{\D_\hI}^\hJ.
\\
\frac{1}{32}~(\CC\Gamma^{(\hI)})_{{\CA}{\CB}}~
(\Gamma_{(\hI)}\CC^{-1})^{{\CC}{\CD}}&=&
\frac12~{\D_{\CA}}^{({\CC}}~{\D_{\CB}}^{{\CD})},
\eea
or equivalently
\bea
\frac{1}{32}~(\CC\Gamma^{(\hI)})_{{\CA}{\CB}}~(\CC\Gamma_{(\hI)})_{{\CC}{\CD}}
&=&
\frac12~\CC_{{\CC}({\CA}}~\CC_{{\CB}){\CD}}.
\label{complet}
\eea

We expand the bosonic generators ${\CZ}_{{\CA}{\CB}}$ using 
$\CC\Gam^{(\hI)}$ as 
\bea
{\CZ}_{{\CA}{\CB}}&=&a \sum_{\hI=1}^{528}~(\CC\Gamma^{(\hI)})_{{\CA}{\CB}}~
{Z}_{(\hI)},~~~~~
{Z}_{(\hI)}~=~\frac{1}{32a}~(\Gamma_{(\hI)}\CC^{-1})^{{\CB}{\CA}}~
{\CZ}_{{\CA}{\CB}},
\eea
where $a$ is a constant.
{}From \bref{Zdagger} ${{Z_{(\hI)}}}$'s
are anti-hermitic for real 
choice of $a$,
\bea
{Z}_{(\hI)}^\dagger&=&-~{Z}_{(\hI)},~~~~~~~~~a^*~=~a.
\label{ZIdagger}
\eea
In terms of ${Z}_{(\hI)}$ the  \osp algebra \bref{ospqq}-\bref{ospzz}
is expressed as
\bea
\{\CQ_{\CA},\CQ_{\CB}\}&=&a~(\CC \Gam^{(\hI)})_{{\CA}{\CB}}~Z_{(\hI)},
\label{ospqq2}\\
{[}\CQ_{\CA},Z_{(\hI)}]&=&\frac{1}{2}(\CQ\Gam_{(\hI)})_{\CA},
\label{ospqz2}\\
{[}Z_{(\hI)},Z_{(\hJ)}]&=&{f_{\hI\hJ}}^\hK Z_{(\hK)},
\label{ospzz2}
\eea
where ${f_{\hI\hJ}}^\hK$~ is the structure constants of $sp(32)$,
\bea
{f_{\hI\hJ}}^\hK~\equiv~-\frac{1}{32}tr(\Gam_{(\hI)}\Gam_{(\hJ)}\Gam^{(\hK)})
\label{spsc}
\eea
$s=-8{a}$ is a normalization convention.
In this form the  \osp algebra the Jacobi identities are 
guaranteed by \bref{complet}
\bea
\sum_{cyclic~{\CB}C{\CD}}~(\CC \Gam^{(\hI)})_{{\CA}{\CB}}(\CC 
\Gam_{(\hI)})_{{\CC}{\CD}}&=&0,
\label{qqqjacobi}
\eea
and the Jacobi identity of ${f_{\hI\hJ}}^\hK$ as
\bea
\sum_{cyclic~\hI\hJ\hK}~{f_{\hI\hJ}}^{\hat L}{f_{\hat L\hK}}^{\hat M}
&=&0,~~~~~~{f_{\hI\hJ}}^{\hat K}~=~-{f_{\hJ\hI}}^{\hat K},
\label{antif}
\\
{[}\frac{1}{2}\Gam_{(\hI)},\frac{1}{2}\Gam_{(\hJ)}]&=&-~{f_{\hI\hJ}}^\hK
\frac{1}{2}\Gam_{(\hK)}.
\label{qqzjacobi}
\eea

\section{ Super \ads }\par
\indent

The bosonic part of the  \osp algebra is $sp(32)$ and contains  
a set of bosonic generators forming a subalgebra,
$so(4,2)\times so(6) \subset sp(32)$. 
The  $SO(4,2)$ is generated by
\bea
Z_{ab},~~Z_{ \hz ,a},~~~(a,b=0,1,2,3,4), 
\eea
where $Z_{ \hz ,a}$ is $j=0$ element of $Z_{j a}$. The algebra is 
\bea
{[}Z_{ \hz ,a},Z_{ \hz ,b}]&=&Z_{ab},~~~~~~~
{[}Z_{ \hz ,a},Z_{cd}]~=~\h_{a[c}Z_{ \hz ,d]},~~~~~~~
\nn\\
{[}Z_{ab},Z_{cd}]&=&
\h_{ad}Z_{bc}-\h_{bd}Z_{ac}-\h_{ac}Z_{bd}+\h_{bc}Z_{ad}.
\label{ppads}\eea
The $SO(6)$ is generated by
\bea
Z_{a'b'},~~Z_{ \hz ,a'},~~~(a',b'=1',2',3',4',5'), 
\eea
as
\bea
{[}Z_{ \hz ,a'},Z_{ \hz ,b'}]&=&-Z_{a'b'},~~~~~~~
{[}Z_{ \hz ,a'},Z_{c'd'}]~=~\h_{a'[c'}Z_{ \hz ,d']},~~~~~~~
\nn\\
{[}Z_{a'b'},Z_{c'd'}]&=&
\h_{a'd'}Z_{b'c'}-\h_{b'd'}Z_{a'c'}-\h_{a'c'}Z_{b'd'}+\h_{b'c'}Z_{a'd'}.
\label{pps5}
\eea
Note the sign difference of \bref{ppads} and \bref{pps5} so that their
algebras are so(4,2) and so(6) respectively. 
They are identified with the \ads generators and the algebra as
\bea
Z_{ \hz ,a}~=~ P_a,~~~~~Z_{ab}~=~ M_{ab}~&,&~
Z_{ \hz ,a'}~=~ P_{a'},~~~~~Z_{a'b'}~=~ M_{a'b'},
\\
\left[P_a,P_b\right]=M_{ab}&,&\left[P_{a'},P_{b'}\right]=-M_{a'b'},\nn\\
\left[P_a,M_{bc}\right]=\eta_{ab}P_c-\eta_{ac}P_b&,&
\left[P_{a'},M_{b'c'}\right]=\eta_{a'b'}P_{c'}-\eta_{a'c'}P_{b'}, \nn\\
\left[M_{ab},M_{cd}\right]=\eta_{bc}M_{ad}+ 3{~\rm terms}&,&
\left[M_{a'b'},M_{c'd'}\right]=\eta_{b'c'}M_{a'd'}+ 3{~\rm terms},
\label{bosonads}\eea
where $~P_a~$ and $~P_{a'}~$ are AdS$_5$ and S$^5$ momenta respectively.
The \ads is a subalgebra of the sp(32) thus of \osp. 

The supersymmetric \ads algebra is the su(2,2$|$4)
and has been extensively discussed in the superstring context. 
It has the same supercharge contents as those of \osp. 
As the \osp is maximally
generalized superalgebra the anti-commutator of two 
supercharges contains all bosonic charges as \bref{ospqq2}
while that of  \sads contains only bosonic \ads charges. 
In this sense \sads cannot be a subalgebra of \osp. 
In this section we examine if the \sads algebra is 
obtained by a ``restriction" of the \osp algebra or not.\footnote{
The ``restriction" means 
1) dropping the commutation relations for the extra generators
$Z_{(I'')}$, 
$[*,Z_{(I'')}]=...$, and
2) dropping the extra generators appearing in the r.h.s. of the 
(anti-)commutators. }
In other words if the \osp is any generalized superalgebra associated 
with the \sads.

The $\CQ\CZ$ commutators of the  \osp algebra are given in \bref{ospqz2} 
and those for \ads generators are read as
\bea
{[}{Q}_{\A\A'A},Z_{ \hz ,a}]~=~\frac12({Q}\gam_{a}(-i\rho_0))_{\A\A'A}
&,~~~&
{[}{Q}_{\A\A'A},Z_{ \hz ,a'}]~=~\frac12({Q}\gam_{a'}\rho_0)_{\A\A'A},
\nn\\
{[}{Q}_{\A\A'A},Z_{ab}]~=~\frac12({Q}\gam_{ba})_{\A\A'A}&,~~~&
{[}{Q}_{\A\A'A},Z_{a'b'}]~=~\frac12({Q}\gam_{b'a'})_{\A\A'A}.
\eea
They show the covariance  of ${Q}$ under the bosonic \ads transformations,
\bea
\left[Q_A,P_a\right]=\frac{i}{2}Q_{B}\gamma_{a}\epsilon_{BA}&,&
\left[Q_{A},P_{a'}\right]=-\frac{1}{2}Q_B\gamma_{a'}\epsilon_{BA},
\nn\\
\left[Q_A,M_{ab}\right]=-\frac{1}{2}Q_{A}\gamma_{ab}&,&
\left[Q_{A},M_{a'b'}\right]=-\frac{1}{2}Q_A\gamma_{a'b'},
\eea
where $\ep=i\tau_2$. 
The explicit form of the  \osp ${\CQ}{\CQ}$ algebra is
\bea
\{{Q}_{\A\A'A},{Q}_{\B\B'B}\}&=&a~\frac12c_{AB}[
(C\gam^{ab})_{\A\B}(C')_{\A'\B'}~Z_{ab}~+~
(C)_{\A\B}(C'\gam^{a'b'})_{\A'\B'}~Z_{a'b'}~+~
\nn\\&&+~
(C\gam^{ab})_{\A\B}(C'\gam^{c'})_{\A'\B'}~Z_{abc'}~+~i~
(C\gam^a)_{\A\B}(C'\gam^{b'c'})_{\A'\B'}~Z_{ab'c'}]
\nn\\&+&a~
(c\rho^j)_{AB}[~(C)_{\A\B}(C')_{\A'\B'}~Z_{j}
\nn\\&&+~i~
(C\gam^{a})_{\A\B}(C')_{\A'\B'}~Z_{ja}~+~
      (C)_{\A\B}(C'\gam^{a'})_{\A'\B'}~Z_{ja'}
\nn\\&&+~i~
(C\gam^{a})_{\A\B}(C'\gam^{b'})_{\A'\B'}~Z_{jab'}~+~
\frac{1}4(C\gam^{ab})_{\A\B}(C'\gam^{c'd'})_{\A'\B'}~Z_{jabc'd'}]
\nn\\
&=&a~ \D_{AB}[-i
(C\gam^{a})_{\A\B}(C')_{\A'\B'}~P_{a}~-~
      (C)_{\A\B}(C'\gam^{a'})_{\A'\B'}~P_{a'}]
\nn\\&+&\frac{a}2~\ep_{AB}[
(C\gam^{ab})_{\A\B}(C')_{\A'\B'}~M_{ab}~+~
(C)_{\A\B}(C'\gam^{a'b'})_{\A'\B'}~M_{a'b'}]
\nn\\&+&....
\label{QQOSP}
\eea
where .... terms are the  \osp generators that are not included in the \ads.
It is compared with that of the \sads algebra
\bea
\left\{Q_{\alpha\alpha'A},Q_{\beta\beta'B}\right\}&=&
2\delta_{AB}\left[
-i{C'}_{\alpha'\beta'}(C\gamma^a)_{\alpha\beta}P_a
+C_{\alpha\beta}(C'\gamma^{a'})_{\alpha'\beta'}P_{a'}
\right]\nn\\
&+&
\epsilon_{AB}\left[  
{C'}_{\alpha'\beta'}(C\gamma^{ab})_{\alpha\beta}M_{ab}
-C_{\alpha\beta}(C'\gamma^{a'b'})_{\alpha'\beta'}M_{a'b'}
\right].
\label{QQADS}
\eea
The \ads generators in the  \osp algebra \bref{QQOSP} and 
the super-AdS$_5\times$S$^5$ algebra
in \bref{QQADS} have different signs.
If we adjust the value $a=2$ so that 
the coefficients of $P$ and $M$ in \bref{QQOSP} and \bref{QQADS} 
coincide, those of  $P'$ and $M'$  have
opposite signs , and {\it vise versa} for $a=-2$.

The sign difference is required by the closure of the algebra.
The ($\CQ\CQ\CQ$) Jacobi identity of \bref{QQOSP}
holds due to the presence of all \osp generators in the r.h.s. of the 
\bref{QQOSP} using with \bref{qqqjacobi} obtained from the completeness
relation \bref{complet}
\bea
\sum_{cyclic~{\CB}{\CC}{\CD}}~\sum_{all~\hI}
(\CC\Gamma^{(\hI)})_{{\CA}{\CB}}~(\CC\Gamma_{(\hI)})_{{\CC}D}&=&0.
\label{cyclosp}
\eea
If we would drop .... terms 
in the \bref{QQOSP} the Jacobi identity no more holds 
\bea
\sum_{cyclic~{\CB}{\CC}{\CD}}~\sum_{I\in AdS_5\times S^5}
(\CC\Gamma^{(I)})_{{\CA}{\CB}}~(\CC\Gamma_{(I)})_{{\CC}D}&\neq &0.
\label{cyclospnon}
\eea
In \bref{QQOSP} the relative signs of $P'$ and $M'$  terms and  
$P$ and $M$ terms are different from those of \bref{QQADS}.
It turns the Jacobi identity for \bref{QQADS} holds using an identity 
independent of \bref{cyclosp};
\bea
\sum_{cyclic~{\CB}{\CC}{\CD}}&&\left(
(\frac12 (\CC\Gam^{ab})_{{\CA}{\CB}}(\CC\Gam_{ab})_{{\CC}{\CD}}+
(\CC\Gam^{\hz a})_{{\CA}{\CB}}(\CC\Gam_{\hz a})_{{\CC}{\CD}})\right.
\nn\\&&~-~\left.
(\frac12 (\CC\Gam^{a'b'})_{{\CA}{\CB}}(\CC\Gam_{a'b'})_{{\CC}{\CD}}+
(\CC\Gam^{\hz a'})_{{\CA}{\CB}}(\CC\Gam_{\hz a'})_{{\CC}{\CD}})~\right)~=~0.
\label{cycladsJ}
\eea
It is proved by using the Fierz identities, which are valid both for
$\gam_a$ and $\gam_{a'}$, 
\bea
\frac12(C\gam^{ab})_{\A\B}(C\gam_{ab})_{\gam\D}&=&
2~(C)_{(\A\gam}(C)_{\B)\D},
\nn\\
(C\gam^{a})_{\A\B}(C\gam_{a})_{\gam\D}&=&2~(C)_{[\A\gam}(C)_{\B]\D}~-~
C_{\A\B}~C_{\gam\D}.
\label{Fierz}
\eea

In summary we have shown that the supersymmetric \ads algebra is {not} 
a restriction of the  \osp and maximal generalization of 
the \sads algebra, if any, is not the  \osp superalgebra. 
In appendix B we obtain the same result for the superconformal algebra
in 4 dimensions as it is also isomorphic to the su(2,2$|$4).

\section{Extensions of the super-AdS$_5\times$ S$^5$ algebra}\par
\indent

In the last section we have shown the \sads algebra is not 
a restriction of the  \osp.
In this section we will find possible generalization 
of the \sads. 
``Generalization" 
here means that the \ads algebra is obtained by 
a restriction of the generators to those of \ads in the 
generalized algebras.
In the generalized algebras we keep the covariance under \ads. 
When we consider some (brane) solutions in the \sads background 
they break also manifest $AdS_5$ and/or $S^5$ covariance.
However it is a spontaneous symmetry breaking
in the presence of a particular solution. 
The \ads covariance of the theory would be still maintained. 
Thus
we impose the assumption that the bosonic AdS$_5\times S^5$ algebra
is a subalgebra of the bosonic part of the superalgebra with brane charges.

Now we try to add bosonic generators on the \sads algebra so that the
commutators satisfy the \Jis. 
We classify possible forms of the bosonic generators in the basis 
of \bref{Gsymgam} as
\bea
\hI~~~\to\pmatrix{I&:&Z_{(\hz,a)},Z_{(ab)}
                \cr I'&:&Z_{(\hz,a')},Z_{(a'b')}
\cr I''_{(0)}&:&Z_{(\hz)}   
\cr I''_{(1)}&:& Z_{(abc')}, Z_{(ab'c')}, 
Z_{(\hz,ab')}, Z_{(\hz,abc'd')}  
\cr I''_{(2)}&:&Z_{(j)}, Z_{(j,a)}, Z_{(j,a')}, 
Z_{(j,ab')}, Z_{(j,abc'd')},~~~~~(j=1,2)   \cr}
\eea
$Z_{(I)},~(I=(\hat 0,a),(ab))$ are $AdS_5$  generators and
$Z_{(I')},~(I'=(\hat 0,a'),(a'b'))$ are $S^5$  generators,~~
($\hat 0$ means $j=0$).
They have subalgebra structures 
\medskip 
\bea
\matrix{(I+I')&\subset&(I+I'+I''_0)&\subset&(I+I'+I''_0+I''_1)&
\subset&(I+I'+I''_0+I''_1+I''_2)\cr \cr
(AdS_5+S^5)&&(AdS_5+S^5+u(1))&&(gl(16))&
&(sp(32)).\cr}
\nn\\
\label{subalg}\eea
$(I+I'+I''_0+I''_1)$ can also be split into $(I+I'+I''_1)+(I''_0)$,
which is $sl(16)+u(1)$. 

\vs

The bosonic part of algebra is
\bea
{[}Z_{(\hI)},Z_{(\hJ)}]&=&s_{\hI}s_{\hJ}{f_{\hI\hJ}}^\hK~Z_{(\hK)}~
\equiv~
{g_{\hI\hJ}}^\hK~Z_{(\hK)},
\eea
where ${f_{\hI\hJ}}^\hK$~ is the structure constants of $sp(32)$ \bref{spsc}.
Values of $s_{\hI}$'s are listed below and are either $``0"$ or $``1"$ 
depending on which bosonic subalgebra is taken into account
\bea
\matrix{
(AdS_5+S^5)&&(AdS_5+S^5+U(1))&&(gl(16))&
&(sp(32))\cr
s_{I}=1&&s_{I}=1 &&s_{I}=1 &&s_{I}=1\cr
s_{I'}=1&&s_{I'}=1 &&s_{I'}=1 &&s_{I'}=1\cr
s_{I''_0}=0&&s_{I''_0}=1&&s_{I''_0}=1&&s_{I''_0}=1\cr
s_{I''_1}=0&&s_{I''_1}=0&&s_{I''_1}=1&&s_{I''_1}=1\cr
s_{I''_2}=0&&s_{I''_2}=0&&s_{I''_2}=0&&s_{I''_2}=1.\cr
}
\label{subalg2}\eea
$(\CZ\CZ\CZ)$ \Ji is verified since the structure constant ${g_{\hI\hJ}}^\hK$ 
satisfies the Jacobi relation for each subalgebra of \bref{subalg2}
\bea
\sum_{cyclic~\hI\hJ\hK}{g_{\hI\hJ}}^{\hat L}{g_{\hK{\hat L}}}^{\hat M}~=~0.
\label{zzzex}
\eea

$\CQ\CZ$ commutators are
\bea
{[}Q_\CA,Z_{(\hI)}]&=&\frac{s_{\hI}}2(Q\Gam_{(\hI)})_\CA,~~~~~~~
\eea
and $(\CQ\CZ\CZ)$ \Ji requires
\bea
&&{[}Q_\CA,[Z_{(\hI)},Z_{(\hJ)}]]~+~{[}Z_{(\hJ)},[Q_\CA,Z_{(\hI)}]]~+~
{[}Z_{(\hI)},[Z_{(\hJ)},Q_\CA]]
\nn\\&=&
\frac{s_{\hK}}2 (Q\Gam_{(\hK)})_\CA~{g_{\hI\hJ}}^\hK ~-~
\frac{{s_{\hI}}s_{\hJ}}4(Q[\Gam_{(\hJ)},\Gam_{(\hI)}])_\CA
\nn\\&=&
\frac{1}2(s_{\hK}-1) {s_{\hI}}s_{\hJ}
~{f_{\hI\hJ}}^\hK ~(Q\Gam_{(\hK)})_\CA~=~0.
\eea
The subgroup structure guarantees that it vanishes for each subalgebra of 
\bref{subalg2}. For example 
for $gl(16)$, $s_{K''_2}=0$ but ${f_{\hI\hJ}}^{K''_2}=0$ for
$\hI,\hJ\neq I''_2$ for which $s_{\hI}=s_{\hJ}=1$.
Then the \Ji is satisfied.
\vs

{}From the \ads covariance the extension of the $\CQ\CQ$ anti-commutator
of the \sads 
will be  
\bea
{\{}Q_\CA,Q_\CB\}&=&a~\left(
(\CC\Gam^{(I)})_{\CA\CB}~Z_{(I)}~-~(\CC\Gam^{(I')})_{\CA\CB}~Z_{(I')}
~+~a_{I''}(\CC\Gam^{(I'')})_{\CA\CB}~Z_{(I'')}~\right)
\nn\\
&=&a~\sum_{\hI=I,I',I''}~a_{\hI}(\CC\Gam^{(\hI)})_{\CA\CB}~Z_{(\hI)}.
\eea
where the coefficients of $AdS_5$ and $S^5$ 
\bea
a_{I}=1,~~~a_{I'}=-1
\label{aads}
\eea
are fixed from the \sads algebra \bref{QQADS}. As will be clear from 
\bref{qqzcond2}~ 
$a_{I''_1}~(a_{I''_2})$ takes the same value for all generators of 
$Z_{(I''_1)}~(Z_{(I''_2)})$.

The $(\CQ\CQ\CZ)$ \Ji requires 
\bea
&&{\{}Q_\CA,[Q_\CB,Z_{(\hI)}]\}~+~{[}Z_{(\hI)},\{Q_\CA,Q_\CB\}]~-~
{\{}Q_\CB,[Z_{(\hI)},Q_\CA]\}
\nn\\&=&
a~\left( \frac{s_{\hI}a_{\hJ}}{2}(\CC[\Gam^{(\hJ)},\Gam_{(\hI)}])_
{\CA\CB}~Z_{(\hJ)}~+~
a_{\hJ}(\CC\Gam^{(\hJ)})_{\CA\CB}~{g_{\hI\hJ}}^{\hK}Z_{(\hK)}~\right)
\nn\\&=&
a~{s_{\hI}}\left(-a_{\hK}
~+~a_{\hJ}s_{\hJ}\right)~{f_{\hI\hat J}}^{\hat K}~
(\CC\Gam^{(\hat J)})_{\CA\CB}~~Z_{(\hK)}~=~0,
\label{qqzcond}
\eea
where we have used \bref{CGam}. The condition it vanishes is 
\bea
{s_{\hI}}\left(-a_{\hK}~+~a_{\hJ}s_{\hJ}\right)~
{f_{\hI\hat J}}^{\hat K}&=&0
\label{qqzcond2}\eea
and satisfied for two cases,

Case 1,~If $Z_{(\hJ)}$ (with non zero $s_{\hJ}$) 
is mixed with $Z_{(\hK)}$ by any of $Z_{(\hI)}$ (with non zero 

\hskip 13mm $s_{\hI}$), i.e. ${f_{\hI\hat J}}^{\hat K}\neq 0~$ in the 
subalgebra under consideration then $a_{\hK}$ is 

\hskip 13mm  necessarily equal to $a_{\hJ}$.

Case 2,~ If $Z_{(\hJ)}$ (with zero $s_{\hJ}$) 
is mixed with $Z_{(\hK)}$ by any of $Z_{(\hI)}$ (with non zero 
$s_{\hI}$),

\hskip 13mm   i.e. 
${f_{\hI\hat J}}^{\hat K}\neq 0~$ in the subalgebra under consideration
then $a_{\hK}$ is necessarily 

\hskip 13mm   equal to $0$. 

\noindent 
In the following we will explain the values of $a_{\hI}$'s
listed below
\bea
\matrix{
(AdS_5+S^5)&&(AdS_5+S^5+u(1))&&(gl(16))&&(sp(32))\cr
s_{I}=1&&s_{I}=1 &&s_{I}=1 &&s_{I}=1\cr
s_{I'}=1&&s_{I'}=1 &&s_{I'}=1 &&s_{I'}=1\cr
s_{I''_0}=0&&s_{I''_0}=1&&s_{I''_0}=1&&s_{I''_0}=1\cr
s_{I''_1}=0&&s_{I''_1}=0&&s_{I''_1}=1&&s_{I''_1}=1\cr
s_{I''_2}=0&&s_{I''_2}=0&&s_{I''_2}=0&&s_{I''_1}=1\cr
a_{I}=1&&a_{I}=1 &&a_{I}=1 &&a_{I}=1\cr
a_{I'}=-1&&a_{I'}=-1 &&a_{I'}=-1 &&a_{I'}=-1\cr
a_{I''_0}=any&&a_{I''_0}={any}&&a_{I''_0}={any}&&a_{I''_0}={any}\cr
a_{I''_1}=0&&a_{I''_1}=0&&a_{I''_1}=\times&&a_{I''_1}=\times\cr
a_{I''_2}=0&&a_{I''_2}=0&&a_{I''_2}=\times&&a_{I''_1}=\times\cr
}
\label{subalg3}\eea
In the first \ads algebra $\hI=I,I'$ and $\hJ=J,J'$ 
are case 1 ($s_\hJ=1$). Since $AdS_5$ generators   commute with the
$S^5$ ones, 
$a_{I}\neq a_{I'}$ is not a contradiction.
$\hI=I,I'$ and $\hJ=J''_0,J''_1,J''_2$  are
case 2 ($s_\hJ=0$). Since $Z_{\hz}$ commutes with \ads 
there appears no condition on $a_{I''_0}$.
$a_{I''_1}=0$ comes, for example, from the following commutators,
\bea
{[}Z_{\hz a},Z_{bcd'}]&=&\h_{a[b}Z_{\hz,c]d'}.
\eea
Since $Z_{\hz a}\in I,~(s_{\hI}=1)$ ~$Z_{bcd'},Z_{\hz,cd'}\in {I''_1},~
(s_{\hJ}=0)$ and ${f_{\hI\hat J}}^{\hat K}\neq 0$ 
then $a_{\hK}=a_{I''_1}=0$. 
Similarly $a_{I''_2}=0$ comes from the following commutators,
\bea
{[}Z_{\hz a},Z_{\hat 1,b}]&=&\h_{ab}Z_{\hat 2}.
\eea
Since $Z_{\hz a}\in I,~(s_{\hI}=1)$ ~$Z_{\hat 1,b},Z_{\hat 2}\in {I''_2}
,~(s_{\hJ}=0)$ and ${f_{\hI\hat J}}^{\hat K}\neq 0$ 
then $a_{\hK}=a_{I''_2}=0$. 

In the second \ads+$u(1)$ algebra $\hI=I,I',I''_0$ and $\hJ=J,J',J''_0$
 are case 1 ($s_\hJ=1$). Since $AdS_5$ and $S^5$  is commuting
it is not necessary to be $a_{I}=a_{I'}$.
Since \ads and $u(1)$ is commuting
there appears no condition on $a_{I''_0}$.
$\hI=I,I',I''_0$ and $\hJ=J''_1,J''_2$  are case 2 ($s_\hJ=0$).
$a_{I''_1}=a_{I''_2}=0$ comes from the same reasons as above.

For the $gl(16)$ algebra there is no consistent solution.
In order to see it, it is sufficient to observe 
\bea
{[}Z_{abc'},Z_{\hz,d}]&=&\h_{[bd}Z_{\hz,a]c'},
\eea
for which $\hI=(abc')\in I''_1,(s_\hI=1),~\hJ=(\hz d)\in I,(s_\hJ=1),~
\hK=(\hz ac')\in I''_1,(s_\hK=1)$ then 
\bea
a_{I''_1}&=&a_{I}~=~1.
\label{condadd1}\eea
On the other hand 
\bea
{[}Z_{ab'c'},Z_{\hz,d'}]&=&\h_{[c'd'}Z_{\hz,ab']},
\eea
for which $\hI=(ab'c')\in I''_1,(s_\hI=1),~\hJ=(\hz d')\in I',(s_\hJ=1),~
\hK=(\hz ac')\in I''_1,(s_\hK=1)$ then 
\bea
a_{I''_1}&=&a_{I'}~=~-1.
\label{condadd2}
\eea
\bref{condadd1} contradicts with \bref{condadd2} due to the opposite
signs of $a_{I}=1$ and $a_{I'}=-1$ \bref{aads}. 

The same argument can be applied to conclude that there is no consistent
solution in $sp(32)$ case. It is consistent with the result in the previous
section that
the \sads algebra is not a restriction of the \osp.
That is the generalization of \ads is failed when the 
$\CQ\CQ$ anti-commutator includes generators non-commuting 
both with $AdS_5$ and $S^5$ generators. Values of $a_{\hI}$
can differ only in different compact subalgebras. 
In the $gl(16)$ and $sp(32)$,~ $a_{I}$ and $a_{I'}$ are 
necessarily to have the same value in order to be consistent 
superalgebras and cannot be accommodated with
\bref{aads}.
\vs

There remains to check the $(\CQ\CQ\CQ)$ Jacobi identity. 
It requires
\bea
&&[Q_\CA,\{Q_\CB,Q_{\CC}\}]~+~[Q_\CB,\{Q_\CC,Q_{\CA}\}]~+~
[Q_\CC,\{Q_\CA,Q_{\CB}\}]
\nn\\&=&
\frac12\sum_{cyclic~\CA\CB\CC}
(Q\CC^{-1})^\CD(\CC\Gam_{(\hJ)})_{\CD\CA}~a_{\hJ}s_{\hJ}~
               (\CC\Gam^{(\hJ)})_{\CB\CC}~=~0
\eea
then
\bea
\sum_{\CA\CB\CC,cyclic}~a_{\hJ}s_{\hJ}~
(\CC\Gam_{(\hJ)})_{\CD\CA}~(\CC\Gam^{(\hJ)})_{\CB\CC}&=&0.
\label{qqqex}
\eea
For \ads case ($s_{\hI''_0}=0)$, it is
\bea
\sum_{\CA\CB\CC,cyclic}~\left(~
(\CC\Gam_{(I)})_{\CD\CA}~(\CC\Gam^{(I)})_{\CB\CC}~-~
(\CC\Gam_{(I')})_{\CD\CA}~(\CC\Gam^{(I')})_{\CB\CC}\right)&=&0
\eea
and is satisfied by the \ads Fierz identity \bref{cycladsJ}.

For \ads+u(1) case ($s_{\hI''_0}=1)$, there appears an additional
term 
\bea
\sum_{\CA\CB\CC,cyclic}~\left(~
(\CC\Gam_{(I)})_{\CD\CA}~(\CC\Gam^{(I)})_{\CB\CC}~-~
(\CC\Gam_{(I')})_{\CD\CA}~(\CC\Gam^{(I')})_{\CB\CC}~+~a_{\hz}
(\CC\Gam_{(\hz)})_{\CD\CA}~(\CC\Gam^{(\hz)})_{\CB\CC}
\right)
\nn\\
\eea
and vanishes only for $a_{\hz}=0$ case. 
\vs

In summary under the present assumption that the
\ads is a bosonic subalgebra of 
bosonic part of the {generalized} algebra only two types of u(1) extension 
are allowed. One is a central extension
\bea
{[}Q,Z_{\hz}]&=&0,~~~
\nn\\
{\{}Q_\CA,Q_\CB\}&=&a~\left(
(\CC\Gam^{(I)})_{\CA\CB}~Z_{(I)}~-~(\CC\Gam^{(I')})_{\CA\CB}~Z_{(I')}
~+~a_{\hz}(\CC\Gam^{(\hz)})_{\CA\CB}~Z_{(\hz)}~\right)
\label{adsext1}\eea
and the other is
\bea
{[}Q,Z_{\hz}]&=&\frac12Q\Gam_\hz,~~~
\nn\\
{\{}Q_\CA,Q_\CB\}&=&a~\left(
(\CC\Gam^{(I)})_{\CA\CB}~Z_{(I)}~-~(\CC\Gam^{(I')})_{\CA\CB}~Z_{(I')}
\right),
\label{adsext2}\eea
in which $Z_{\hz}$ is not an ideal of the algebra.

\section{Summary and Discussions }\par
\indent

We have shown that the su(2,2$|$4), the super-AdS$_5\times S^5$ algebra or 
the four-dimensional $\cal N$=4 superconformal algebra, 
is not a restriction of the osp(1$|$32).
The situation is the same for the super-pp-wave algebra as it is obtained by
the Penrose limit of the super-AdS$_5\times S^5$ algebra\cite{BFP,HKS}. 
In addition, we have shown that
under the assumption that bosonic AdS$_5\times S^5$ algebra is a subalgebra of
the bosonic part of the superalgebra with brane charges,
only two trivial types of u(1) extension of the super-AdS$_5\times S^5$
algebra are allowed.

Our results suggest that
the {generalized} superalgebra associated with branes
in the AdS$_5\times S^5$ background, if any,
cannot be any {generalization}
of the super-AdS$_5\times S^5$ algebra which contains 
the super-AdS$_5\times S^5$ algebra as a restriction.
This is completely different from the flat case,
where extended superalgebra with brane charges contains
the super-Poincare algebra as a restriction.
The property that brane charge affects the algebra associated
with the background may be related to a back reaction of the brane
on the background. 
If this is the case, the brane probe analysis on the extended superalgebra
must be modified appropriately.

There are possible ways to obtain {generalized} superalgebras
with brane charges corresponding to
branes in the AdS$_5\times S^5$ background.
First,
one examines all possible commutation relations among two of Noether charges.
Secondly,
one examines small fluctuations around a brane solution in the 
AdS$_5\times S^5$
background of supergravity.
Such a solution is not known well yet.
Recently, D-brane supergravity solutions in the pp-wave background
were constructed \cite{BMZ}.
One may construct a {generalized} superalgebra with the brane charge
recovering (super)symmetries broken by the brane.
Thirdly, it is known that an intersection solution
of a stack of D3-branes and a D$p$-brane in flat background
is transformed to a D$p$-brane solution in the AdS$_5\times S^5$ background
under a near-horizon limit.
One expects that the extended super-Poincare algebra
with D3- and D$p$-brane charges
is transformed to a {generalized} superalgebra
with a D$p$-brane charge associated with
a D$p$-brane solution in the AdS$_5\times S^5$ background.
In order to show this,
at first, we must derive the AdS$_5\times S^5$ algebra
from the super Poincare algebra with D3-brane charges.
{We leave these issues for future investigations.}

{\bf Acknowledgements}

 We are grateful to Joaquim Gomis, Machiko Hatsuda, Antoine Van Proeyen and
Joan Sim\'on for useful discussions and comments. 


\appendix

\section{ Uniqueness of the \osp algebra }
\indent

We show that any maximally {generalized} 
supersymmetric algebra 
is expressed in the form of \bref{ospqq}-\bref{ospzz} by redefinitions
explicitly. 
In a maximally {generalized}
 algebra $\frac{N(N+1)}{2}$ independent 
bosonic charges appear in the anti-commutator of the 
$N$-supercharges $\CQ_\CA$, (in the present case $N=32$).   
Then the bosonic charges $\CZ_{\CA\CB}$ are defined by 
the first equation \bref{ospqq} 
\bea
{\CZ}_{{\CA}{\CB}}&\equiv&\{\CQ_{\CA},\CQ_{\CB}\}~
=~{\CZ}_{{\CB}{\CA}}.
\label{ospqq1}
\eea
In the closed algebra 
${[}\CQ_{\CA},{\CZ}_{{\CB}{\CC}}]$ is odd and generally written as
\bea
{[}\CQ_{\CA},{\CZ}_{{\CB}{\CC}}]&=&{[}\CQ_{\CA},\{\CQ_{\CB},\CQ_{\CC}\}]~
=~\CQ_{\CE}\W^{\CE}_{{\CA},{\CB}{\CC}},~~~~~
\W^{\CE}_{{\CA},{\CB}{\CC}}~=~\W^{\CE}_{{\CA},{\CC}{\CB}}.
\label{ospqz1}
\eea
The $(\CQ\CQ\CQ)$ Jacobi identity 
\bea
\sum_{cyclic~{\CA}{\CB}{\CC}}[\{\CQ_\CA,\CQ_\CB\},\CQ_\CC]~=~0,~~~~
\eea
requires
\bea
\sum_{cyclic~{\CA}{\CB}{\CC}}\W^{\CE}_{{\CA},{\CB}{\CC}}&=&0.
\label{cyclic}
\eea
The $(\CQ\CQ\CZ)$ Jacobi  identity requires
\bea
{[}{\CZ}_{{\CA}{\CB}},{\CZ}_{{\CC}{\CD}}]&=&{\CZ}_{({\CA}{\CE}}
\W^{\CE}_{{\CB}),{{\CC}{\CD}}}~=~
-~{\CZ}_{({\CC}{\CE}}\W^{\CE}_{{\CD}),{{\CA}{\CB}}}.
\label{ospzz1} 
\eea
The last equality comes from anti-symmetry under 
$({\CA}{\CB}\leftrightarrow {\CC}{\CD})$. It follows 
\bea
\D^{{\CF}}_{({\CA}} \W^{{\CG}}_{{\CB}),{{\CC}{\CD}}}~+~\D^{{\CF}}_{({\CC}} 
\W^{{\CG}}_{{\CD}),{{\CA}{\CB}}}~+~
({\CF}\leftrightarrow {\CG})~=~0.
\label{omega1} 
\eea
By contraction with $\D_{\CF}^{\CA}$
\bea
(N+1)~\W^{{\CG}}_{{\CB},{{\CC}{\CD}}}~+~
\D^{{\CG}}_{{\CB}} \W^{{\CF}}_{{\CF},{{\CC}{\CD}}}~+~
\D^{{\CG}}_{({\CC}} \W^{{\CF}}_{{\CD}),{{\CF}{\CB}}}~=~0.
\label{eq2}
\eea
By further contraction with $\D_{\CG}^{\CB}$ we find
\bea
\W^{{\CG}}_{{\CG},{{\CC}{\CD}}}~=~0
\eea
and then
\bea
\W^{{\CG}}_{{\CB},{{\CC}{\CD}}}&=&
\D^{{\CG}}_{({\CC}} \W_{{\CB}{\CD})},~~~~~
\W_{{\CB}{\CD}}\equiv-\frac{1}{(N+1)} \W^{{\CF}}_{{\CD},{{\CF}{\CB}}}.
\label{eq3}
\eea
The cyclic condition \bref{cyclic} means anti-symmetry of 
$\W_{{\CA}{\CB}}$ and \bref{ospqz1} return to the 
form of \bref{ospqz}
\bea
{[}\CQ_{\CA},{\CZ}_{{\CB}{\CC}}]&=&
\CQ_{({\CB}}\W_{{\CA}{\CC})}.
\label{ospqz3}
\eea
The bosonic algebra \bref{ospzz1} becomes symplectic form, $sp(N)$, 
\bea
{[}{\CZ}_{{\CA}{\CB}},{\CZ}_{{\CC}{\CD}}]&=&
{\CZ}_{{(\CA}{\CE}}\W^{\CE}_{{\CB)},{{\CC}{\CD}}}~=~
{\CZ}_{{\CA}({\CC}}\W_{{\CB}{\CD})}~+~{\CZ}_{{\CB}({\CC}}\W_{{\CA}{\CD})}
\label{ospzz2}
\eea
which coincide with \bref{ospzz}. 
All the Jacobi identities hold identically. 

In getting \bref{omega1} we have assumed that all bosonic generators
${\CZ}_{{\CA}{\CB}}$ appear in the right hand of the 
${{\CQ}{\CQ}}$ anti-commutator \bref{ospqq1}. 
For a non-maximally {generalized} superalgebra  
${\CZ}_{{\CA}{\CB}}$'s are not independent and
\bref{omega1} no more holds.
The Jacobi identity is not trivially satisfied 
but is examined case by case. 
\vs

\section{osp(1$|$32) and four-dimensional $\cal N$=4 superconformal 
algebra}
\noindent

In this appendix we show that  
the superconformal algebra $su(2,2|4)$ is also not a restriction of the 
\osp and there is no maximally {generalized} superalgebra. 
It is based on one to one correspondence between generators of the 
\sads and \scn algebras. 

We explicitly write down relations among generators of the
 \osp, the \sads and the \scn algebras based on the isomorphism
\bea
AdS_5&\sim& SO(4,2)~\sim~4D~{\rm conformal},~~~~S^5~\sim~SO(6)~\sim~SU(4) 
\eea        
and their supercharges.

$AdS_5$ generators can be identified to the SO(4,2) generators by 
\bea
P_a ~=~Z_{\hz a}&\equiv& M_{a\sharp},~~~~~~Z_{ab}~\equiv~M_{ab}.
\eea
The so(4,2) algebra is
\bea
\left[M_{AB},M_{CD}\right]&=&\eta_{BC}M_{AD}+ 3{~\rm terms},
\label{bosonso42}
\eea
where $~
A,B=01234\sharp,$ and $\h_{AB}~=~(-;++++;-).$
The sign of the metric in $\sharp$ direction comes from 
\bea
\left[M_{a\sharp},M_{b\sharp}\right]~=~
M_{ab}~=~-\h_{\sharp\sharp}~M_{ab}.
\eea
The conformal generators in 4 dimensions can be composed as 
\bea
\t P_\mu &\equiv& M_{\mu 4}+M_{\mu \sharp}~=~Z_{\mu 4}+Z_{\hz \mu},~~~~~~
M_{\mu\nu}~\equiv~ Z_{\mu\nu},
\nn\\ 
K_\mu &\equiv& M_{\mu 4}-M_{\mu \sharp}~=~Z_{\mu 4}-Z_{\hz \mu},~~~~~~
D ~=~M_{4 \sharp}~=~Z_{\hz 4},
\nn\\ &&\mu=0123,~~~~~~\h_{\mu\nu}~=~(-;+++).
\eea
They satisfy
\bea
\left[\t P_{\mu},\t P_{\nu}\right]&=&\left[K_{\mu},K_{\nu}\right]~=~0
,~~~~
\left[\t P_{\mu},K_{\nu}\right]~=~-2 ~M_{\mu\nu}~+~2~\h_{\mu\nu}~D,
\nn\\
\left[\t P_{\mu},M_{\rho\s}\right]&=&\eta_{\mu[\rho}~\t P_{\s]},~~~~~~~~
\left[K_{\mu},M_{\rho\s}\right]~=~\eta_{\mu[\rho}~K_{\s]},
\nn\\
\left[\t P_{\mu},D\right]&=&\t P_{\mu},~~~~~~~~~~~~
\left[K_{\mu},D \right]~=~ -K_{\mu},
\label{bosonconf}\eea

$S^5$ generators $M_{a'b'}$ and $P_{a'}$ become the SO(6) generators by
\bea
P_{a'}~=~Z_{\hz a'}&\equiv&M_{a'\flat},~~~~~~Z_{a'b'}~\equiv~M_{a'b'}, 
\\
\left[M_{A'B'},M_{C'D'}\right]&=&\eta_{B'C'}M_{A'D'}+ 3{~\rm terms},
\label{bosonso6}\eea
where  
$A',B'=1'2'3'4'5'\flat,~$ and $~\h_{A'B'}~=~(+++++;+).$
SU(4) generators are constructed from $M_{A'B'}$ as 
\bea
{U^{\gam'}}_{\D'}&\equiv&
\frac14~M_{a'b'}~{(\gam^{a'b'})^{\gam'}}_{\D'}~+~
\frac{i}2~M_{a'\flat}~{(\gam^{a'})^{\gam'}}_{\D'}.
\label{defU}
\eea
It is traceless and hermitic and contains 15 independent components. 
Using the Fierz identities \bref{Fierz} 
it is inverted as
\bea
M_{a'b'}&=&\frac12~tr~(U\gam_{b'a'}),~~~~~~
M_{a'\flat}~=~-\frac{i}2~tr~(U\gam_{a'}).
\eea
The su(4) 
 algebra follows from that of so(6) \bref{bosonso6} as 
\bea
{[}{U^{\A'}}_{\B'},{U^{\gam'}}_{\D'}]&=& 
{U^{\A'}}_{\D'}~{\D^{\gam'}}_{\B'}~-~{U^{\gam'}}_{\B'}~{\D^{\A'}}_{\D'}.
\label{SU4alg}
\eea

In the conformal group the generators are graded by their conformal 
dimensions.  
\bref{bosonconf} shows
$\hat P_\mu$ has the conformal dimension $+1$, 
$K_\mu$ has $-1$ and $M_{\mu\nu}$ and $D$ itself have $0$. 
The SU(4) generators $~{U^{\A'}}_{\B'}~$ also have the conformal dimension 
$0$. The 32 supercharges $Q_{\A\A' A}$ are recombined to be 16 
$\bQ$ with the conformal 
dimension $\frac12$ and $\bS$ with $-\frac12$. They are 
\bea
\bQ_{\B i}&=&(Q_{\A\A'1}+iQ_{\A\A'2}){(h_-)^\A}_\B{\D^{\A'}}_i
\nn\\
{\bQ_{\B }}^i&=&(Q_{\A\A'1}-iQ_{\A\A'2}){(h_+)^\A}_\B{C'^{-1}}^{\A'i}
\nn\\
\bS_{\B i}&=&(Q_{\A\A'1}+iQ_{\A\A'2}){(h_+)^\A}_\B{\D^{\A'}}_i
\nn\\
{\bS_{\B}}^i&=&(Q_{\A\A'1}-iQ_{\A\A'2}){(h_-)^\A}_\B{C'^{-1}}^{\A'i}.
\eea
Here $h_\pm$ are chiral projection operators in 4 dimensions 
($\gam^4$ is one usually denoted as $\gam^5$) 
\bea
h_\pm&=&\frac{1\pm \gam^4}{2},~~~h_\pm^\dagger~=~h_\pm.
\eea

They are related by the hermitian conjugation defined by 
\bref{conjugation} as
\bea
\bQ_{\B i}~~\to~~(\bQ^\dagger)^{\B i}&=&
{\bQ_{\gam}}^i~(-\gam^0C^{-1})^{\gam\B}
\nn\\
{\bQ_{\B}}^i~~\to~~ {(\bQ^\dagger)^{\B}}_ i&=&
\bQ_{\gam i}~(\gam^0{C}^{-1})^{{\gam}\B}.
\nn\\
\bS_{\B i}~~\to~~(\bS^\dagger)^{\B i}&=&
{\bS_{\gam}}^i~(-\gam^0C^{-1})^{\gam\B}
\nn\\
{\bS_{\B}}^i~~\to~~ {(\bS^\dagger)^{\B}}_ i&=&
\bS_{\gam i}~(\gam^0{C}^{-1})^{{\gam}\B}.
\label{QSconj}\eea

The $\CQ\CZ$ commutators of the \osp are expressed as follows. 
The commutators with conformal generators are 
\bea
\matrix{
{[}\bQ_{i},M_{\mu\nu}]&=&\frac12 ~\bQ_{i}~\gam_{\nu\mu}&,&
{[}\bQ_{i},D]&=&\frac12 ~\bQ_{i} \cr
{[}\bQ_{i},\hat P_{\mu}]&=&0 &,&
{[}\bQ_{i},K_{\mu}]&=& \bS_{i}~\gam_{\mu}  \cr
{[}\bQ^{i},M_{\mu\nu}]&=&\frac12 ~\bQ^{i}~\gam_{\nu\mu}&,&
{[}\bQ^{i},D]&=&\frac12 ~\bQ^{i}\cr
{[}\bQ^{i},\hat P_{\mu}]&=&0&,&
{[}\bQ^{i},K_{\mu}]&=&-\bS^{i}~\gam_{\mu} \cr
{[}\bS_{i},M_{\mu\nu}]&=&\frac12 ~\bS_{i}~\gam_{\nu\mu}&,&
{[}\bS_{i},D]&=&-~\frac12 ~\bS_{i}\cr
{[}\bS_{i},\hat P_{\mu}]&=&\bQ_{i}~\gam_{\mu}&,&
{[}\bS_{i},K_{\mu}]&=&0 \cr
{[}\bS^{i},M_{\mu\nu}]&=&\frac12 ~\bS^{i}~\gam_{\nu\mu}&,&
{[}\bS^{i},D]&=&-~\frac12 ~\bS^{i}\cr
{[}\bS^{i},\hat P_{\mu}]&=&-~\bQ^{i}~\gam_{\mu}&,&
{[}\bS^{i},K_{\mu}]&=&0,\cr}
\label{bSMcom}\eea
and the commutators with SU(4) generators are obtained from \bref{defU}
\bea \matrix{
\left[\bQ_{\A i},{U^{j}}_{k}\right]&=&
\bQ_{\A k} {\D^{j}}_{i}~-~
\frac{1}4~\bQ_{\A i}~{\D^{j}}_{k},\cr
\left[{\bQ_{\A}}^{i},{U^{j}}_{k}\right]&=&
- {\bQ_{\A}}^{j} {\D^{i}}_{k}~+~
\frac{1}4~{\bQ_{\A}}^{i}~{\D^{j}}_{k},\cr
\left[\bS_{\A i},{U^{j}}_{k}\right]&=&
\bS_{\A k} {\D^{j}}_{i}~-~
\frac{1}4~\bS_{\A i}~{\D^{j}}_{k},\cr
\left[{\bS_{\A}}^{i},{U^{j}}_{k}\right]&=&
- {\bS_{\A}}^{j} {\D^{i}}_{k}~+~
\frac{1}4~{\bS_{\A}}^{i}~{\D^{j}}_{k}.}
\label{QM42}
\eea
There are also commutators of $Q$ with bosonic generators $Z_{(I'')}$ 
which are not in \ads. 

The \osp $\CQ\CQ$ anti-commutators 
in terms of $\bQ$ and $\bS$ are
\bea
\{\bQ_{\A i},\bQ_{\B j}\}&=&
2a[~(Ch_{-})_{\A\B}C'_{ij}(Z_+-iZ_{+4})~+~
(Ch_{-})_{\A\B}~(C'\gam^{a'})_{ij}(Z_{+a'}-{i}Z_{+4a'})
\nn\\
&+&\frac14(C\gam^{\mu\nu}h_{-})_{\A\B}(C'\gam^{a'b'})_{ij}
Z_{+\mu\nu a'b'}~],
\label{bQbQ}\\
\{\bS_{\A i},\bS_{\B j}\}&=&
2a[~(Ch_{+})_{\A\B}C'_{ij}(Z_++iZ_{+4})~+~
(Ch_{+})_{\A\B}~(C'\gam^{a'})_{ij}(Z_{+a'}+{i}Z_{+4a'})
\nn\\
&+&\frac14(C\gam^{\mu\nu}h_{+})_{\A\B}(C'\gam^{a'b'})_{ij}
Z_{+\mu\nu a'b'}~],
\\
\{\bQ_{\A i},\bS_{\B j}\}&=&
2a[~i(C\gam^{\mu}h_{+})_{\A\B}C'_{ij}Z_{+\mu}~+~
{i}(C\gam^{\mu}h_{+})_{\A\B}(C'\gam^{a'})_{ij}Z_{+\mu a'}~
\nn\\&& +~\frac12(C\gam^{\mu}h_{+})_{\A\B}(C'\gam^{a'b'})_{ij}
Z_{+\mu 4a'b'}~],
\\
\{{\bQ_{\A}}^i,{\bQ_{\B}}^j\}&=&
-2a[~
(Ch_{+})_{\A\B}{{C'}^{-1}}^{ij}(Z_-+iZ_{-4})~+~
(Ch_{+})_{\A\B}~(\gam^{a'}{{C'}^{-1}})^{ij}(Z_{-a'}+{i}Z_{-4a'})
\nn\\
&+&\frac14(C\gam^{\mu\nu}h_{+})_{\A\B}(\gam^{a'b'}{{C'}^{-1}})^{ij}
Z_{-\mu\nu a'b'}~],
\\
\{{\bS_{\A}}^i,{\bS_{\B}}^j\}&=&
-2a[~
(Ch_{-})_{\A\B}{{C'}^{-1}}^{ij}(Z_--iZ_{-4})~+~
(Ch_{-})_{\A\B}~(\gam^{a'}{{C'}^{-1}})^{ij}(Z_{-a'}-{i}Z_{-4a'})
\nn\\
&+&\frac14(C\gam^{\mu\nu}h_{-})_{\A\B}(\gam^{a'b'}{{C'}^{-1}})^{ij}
Z_{-\mu\nu a'b'}~],
\\
\{{\bQ_{\A}}^i,{\bS_{\B}}^j\}&=&
-2a[~i(C\gam^\mu h_{-})_{\A\B}{{C'}^{-1}}^{ij}Z_{-\mu}~+~i
(C\gam^\mu h_{-})_{\A\B}~(\gam^{a'}{{C'}^{-1}})^{ij}Z_{-\mu a'}
\nn\\
&-&\frac12(C\gam^{\mu}h_{-})_{\A\B}(\gam^{a'b'}{{C'}^{-1}})^{ij}
Z_{-\mu 4 a'b'}~],
\\
\{{\bQ_{\A}}^i,{\bQ_{\B}}_j\}&=&
2ai[~(C\gam^{\mu}h_{-})_{\A\B} {\D^i}_j ~\hat P_{\mu}
 \nn\\
&+&i
(C\gam^{\mu}h_{-})_{\A\B}
\left(\frac{1}2{(\gam^{a'b'})^i}_j(Z_{\hz\mu 4a'b'}-Z_{\mu a'b'})~-~
i~{(\gam^{a'})^i}_j(Z_{\hz\mu a'}+Z_{\mu 4a'})\right)~],
\nn\\
\\
\{{\bQ_{\A}}^i,{\bS_{\B}}_j\}&=&
2ai[~(Ch_{+})_{\A\B} {\D^i}_j ~D~-~
\frac{1}2
(C\gam^{\mu\nu}h_{+})_{\A\B}{\D^i}_j M_{\mu\nu}~-~2(Ch_{+})_{\A\B}~{U^i}_j
\nn\\&-&i~(Ch_{+})_{\A\B} {\D^i}_j ~Z_{\hz}~+~
(Ch_{+})_{\A\B}~{(\gam^{a'})^i}_j~Z_{\hz 4a'}~-~
\frac{i}4(C\gam^{\mu\nu}h_{+})_{\A\B}{(\gam^{a'b'})^i}_jZ_{\hz\mu\nu a'b'}
\nn\\
&-&\frac{1}{2}~(C\gam^{\mu\nu}h_{+})_{\A\B}{(\gam^{a'})^i}_j ~Z_{\mu\nu a'}~
-~\frac{i}{2}~(Ch_{+})_{\A\B}{(\gam^{a'b'})^i}_j ~Z_{4 a'b'}~],
\label{acQdS}
\\
\{{\bS_{\A}}^i,{\bQ_{\B}}_j\}&=&
2ai[~-(Ch_{-})_{\A\B} {\D^i}_j ~D~-~
\frac{1}2
(C\gam^{\mu\nu}h_{-})_{\A\B}{\D^i}_j M_{\mu\nu}~-~
2(Ch_{-})_{\A\B}~{U^i}_j
\nn\\&-&i~(Ch_{-})_{\A\B} {\D^i}_j ~Z_{\hz}~-~
(Ch_{-})_{\A\B}~{(\gam^{a'})^i}_j~Z_{\hz 4a'}~-~
\frac{i}4(C\gam^{\mu\nu}h_{-})_{\A\B}{(\gam^{a'b'})^i}_jZ_{\hz\mu\nu a'b'}
\nn\\
&-&\frac{1}{2}~(C\gam^{\mu\nu}h_{-})_{\A\B}{(\gam^{a'})^i}_j ~Z_{\mu\nu a'}~
+~\frac{i}{2}~(Ch_{-})_{\A\B}{(\gam^{a'b'})^i}_j ~Z_{4 a'b'}~],
\label{acSdQ}
\\
\{{\bS_{\A}}^i,{\bS_{\B}}_j\}&=&
2ai[~-(C\gam^{\mu}h_{+})_{\A\B} {\D^i}_j~K_{\mu} 
 \nn\\
&-&i(C\gam^{\mu}h_{+})_{\A\B}\left(
\frac{1}2{(\gam^{a'b'})^i}_j(Z_{\hz\mu 4a'b'}+Z_{\mu a'b'})~+i~
{(\gam^{a'})^i}_j(Z_{\hz\mu a'}-Z_{\mu 4a'})\right)].
\label{bSbS}\eea
Here the bosonic generators other than the \scn ones are $Z$'s and
$Z_{\pm*}=Z_{\hat 1*}\pm iZ_{\hat 2*}$. 
These commutators are consistent under the conjugations \bref{QSconj}. 
For example the conjugation of the \bref{acQdS} gives \bref{acSdQ}.
\vs

We have written the \osp algebra in terms of \scn generators and other 
bosonic generators $Z$'s. 
The commutators \bref{bosonconf}, \bref{SU4alg}, \bref{bSMcom}-
\bref{QM42} 
have the same forms as those of the \scn algebra. For the anti-commutators
of the supercharges \bref{bQbQ}-\bref{bSbS},
if we would drop the bosonic generators $Z$'s other than $\hat P_{\mu}, 
K_{\mu}, M_{\mu\nu}, D$ and ${U^i}_j$,
they appear those of the \scn algebra
except that the signs in front of the ${U^i}_j$ terms are reversed. 
The reason is the same as the \ads case. The restriction of the extra generators 
$Z$'s from the \osp is not sufficient to guarantee the Jacobi identities.
In order to do it, it is necessary to reverse the signs of the  ${U^i}_j$ 
terms in the  anti-commutators of the supercharges 
since SU(4) generator  ${U^i}_j$ is related to the $S^5$ generators
by \bref{defU} and \bref{cycladsJ} is applied. 

The {generalization} of the \scn algebra is discussed
{as was done in the case of the \sads in section 4.}
It is not {generalized} up to maximal but only two types of u(1) extension
are allowed. The U(1) generator $Z_{\hz}$ can be combined with the SU(4) 
generator ${U^i}_j$ to form a U(4) generator.


\end{document}